\documentclass[reprint,aps,prd,floats,floatfix,amsmath,amssymb,amsfonts,nofootinbib,longbibliography,superscriptaddress]{revtex4-2}

\usepackage{cases,empheq}
\usepackage{graphicx}
\usepackage{dcolumn}
\usepackage{bm}
\usepackage{natbib}
\usepackage[export]{adjustbox}
\usepackage[hyperindex,colorlinks]{hyperref}
\usepackage{xcolor}

\renewcommand{\vec}{\boldsymbol}

\DeclareFontFamily{U}{mathb}{\hyphenchar\font45}
\DeclareFontShape{U}{mathb}{m}{n}{
      <5> <6> <7> <8> <9> <10> gen * mathb
      <10.95> mathb10 <12> <14.4> <17.28> <20.74> <24.88> mathb12
      }{}

\begin{document}

\title{Perturbative aspects of analogue FLRW spacetime Jellium models}
\author{Matheus E. Pereira}
 \email{mepereira@id.uff.br}
 \affiliation{%
  Instituto de Ci\^encias Exatas, Universidade Federal Fluminense,
  \\ 27213-145 Volta Redonda --- RJ, Brazil
}%
\affiliation{Programa de P\'os Gradua\c{c}\~ao em F\'isica, Instituto de F\'\i sica, Universidade Federal Fluminense,\\
	24210-346 Niter\'oi --- RJ, Brazil}

\author{Hermano Velten}
 \email{hermano.velten@ufop.edu.br}
 \affiliation{Departamento de F\'isica, Instituto de Ciências Exatas e Biológicas, Universidade Federal de Ouro Preto, Ouro Preto - MG, 35402-136, Brasil
}%
\affiliation{Instituto de Astronom\'ia Te\'orica y Experimental, CONICET-UNC, Laprida 854, X5000BGR, C\'ordoba, Argentina}

 \author{Francisco Bento Lustosa}
 \email{chico.lustosa@uece.br} \affiliation{Universidade Estadual do Cear\'a, Faculdade de Educa\c c\~ao, Ci\^encias e Letras de Iguatu, 63500-000, Iguatu, CE, Brazil.}
\author{Alexandre G. M. Schmidt}%
 \email{agmschmidt@id.uff.br}
\affiliation{%
  Instituto de Ci\^encias Exatas, Universidade Federal Fluminense,
  \\ 27213-145 Volta Redonda --- RJ, Brazil
}%
\affiliation{Programa de P\'os Gradua\c{c}\~ao em F\'isica, Instituto de F\'\i sica, Universidade Federal Fluminense,\\
	24210-346 Niter\'oi --- RJ, Brazil}

\date{\today}

\begin{abstract}
We study electro-acoustic perturbative modes in homogeneous, isotropic and expanding --- dubbed as Friedmann-Lemaitre-Robertson-Walker (FLRW) --- Jellium models, thereby mimicking density perturbations in analogous Newtonian cosmological expansions. We present both novel analytic solutions for linear perturbations in specific analogue cosmological expansions and full numerical evaluations that characterize the temporal evolution of the electro-acoustic modes capturing their full dynamical behavior across the linear and the nonlinear regimes. For both the case of pressure supported evolution or modes sourced by nonadiabatic contributions, we also characterize the temporal evolution of such perturbations by introducing their scale dependent particle number fluctuation power spectrum which can act as a tool to connect theory and experiments. The dependence of the latter on the physical parameters of the model is demonstrated in detail.   
\end{abstract}

\maketitle

\section{Introduction}
\label{sec:intro}


\textcolor{black}{The wide range of theoretical and technical developments that have driven modern Physics lead to multiple highly specialized fields that at first glance might seem disconected, from gravitational physics, to condensed matter and high energy particle physics. However, fundamental developments in those fields have been aided by the recognition of similarities and analogies between mathematical structures and/or physical phenomena from different fields. Black hole physics is one of the main areas of gravitational physics to benefit from the theoretical interplay between General Relativity, Quantum Field Theory and Thermodynamics to push forward our understanding of gravitational collapses, singularities, horizons and quantum gravity \cite{carlaalmeida2023analogue}. This theoretical interplay has given rise} to a line of research commonly referred to as Analogue Gravity \cite{analogue-gravity-review}, in which complex and almost inaccessible phenomena can be mimicked in laboratory experiments \cite{jacquet2020next}. \textcolor{black}{From its original motivation to understand the origin of Hawking radiation through analogue experiments \cite{unruh01}, the field of Analogue Gravity has developed into a wider field of research that has provided experimental evidence of emission of radiation from artificial horizons \cite{leonhardt2008,weinfurtner2011, Rubino_2011,Drori2019} but also proposed multiple other experiments that could mimick other gravitational effects that are of interest both for astrophysics \cite{Hartley_gravwaves} and cosmology \cite{barcelo2003analogue,fedichev2004cosmological,lidsey2004cosmic,Weinfurtner2007,prain2010analogue,faraoni2020lagrangian,velten2012power,de2015nonlinear,fifer2019analog,steinhauer2022analogue,niehof2026freezing,faraoni2026laboratory,hyperlente, cosmologia}. In this work we will focus on an analogue cosmological framework employing the Jellium model \cite{BohmPines1951,BohmPines1952,BohmPines1953} exploring  the possibility of mimicking cosmological expansion with an experiment that has been previously suggested in \cite{kolomeisky2019natural, kolomeisky2020normal, kolomeisky2021analog}. The wider consequences and possibilities to test cosmological models and observational effects have not been studied in detail before and we present a first step in this direction here by analyzing the evolution of perturbations for different backgrounds and proposing a path to test the effects of particular choices of cosmological evolutions.}


The Jellium model represents a simple physical system that is conventionally represented by the uniform electron gas (UEG), also known as the homogeneous electron gas (HEG) \cite{BohmPines1953,PinesNozieres1966}. In this framework, electrons move within a continuous, positively charged background that ensures the system remains neutral on average, thereby enabling the study of electron–electron interactions without the complexity of an ionic lattice. In this sense, the Jellium model emerges as a phenomenological approach to representing quantum mechanical systems from the perspective of well‑established macroscopic physics \cite{Wigner1934,Brack1993,Mahan2000}. The Jellium model has demonstrated a remarkable richness of applications, notably within the analogue gravity program. Ref. \cite{kolomeisky2019natural} has studied an expanding Jellium configuration with a background velocity field motivated by the cosmological Hubble flow. Later, in Ref. \cite{kolomeisky2020normal}, this same system has been studied within a first order perturbative formalism and its normal modes of vibrations have been identified. Such perturbative structure resembles the same one found in scalar perturbations of the so called Newtonian cosmology \cite{McCREA1955,Mukhanov:2005sc}. In fact, this investigation on the evolution of inhomogeneities in the Jellium model is justified since the assumption of a uniform electron density no longer holds in finite systems, where the density displays relevant spatial oscillations and decays rapidly near the surface, thereby making the simple uniform-density model inadequate \cite{Krotscheck1986,Cohen2012}.

The goal of the present work is twofold: by revisiting the expanding Jellium model, we provide a different Friedmann-like equation for the background expansion in comparison to the one presented in Ref. \cite{kolomeisky2019natural}. Then, we extend the main result of Ref. \cite{kolomeisky2020normal} by presenting a larger class of solutions based on the adiabatic Jellium expansion approach.     

By performing a physical Jellium model experiment conducted under highly controlled conditions, one can directly test the validation of theory and its features, i.e., it allows experimental verification of predictions from the uniform electron gas model, such as screening, plasmon excitations, and correlation energies. Hence, the analogy between Jellium and cosmology is indeed interesting since the background expansion of the former can be manipulated in the laboratory, contrary to the cosmological case. We therefore open a broader class of possibilities for studying the Maxwell-Poisson system. In order to extend the theoretical analysis of the expanding Jellium model to physically meaningful (and realistic) cosmological scenarios we study different background expansions, providing hence a richer phenomenology to this problem.

A secondary goal is to analyse special cases with analytical and numerical solutions that could, in principle, be verified experimentally, bridging cosmology and plasma physics or condensed matter physics. Indeed, given that cosmology is currently based mostly on observations of electromagnetic signals and numerical simulations, but not experiments, this Jellium toy model may bring a new perspective to current research on clustering properties (see Refs. \cite{koskinen1995electron,Cricchio} for interesting aspects of the problem) by providing a platform for experimental investigations.

In this context, we follow Kolomeisky's approach \cite{kolomeisky2019natural,kolomeisky2020normal,kolomeisky2021analog} to determine the background evolution of 3D expanding Jelllium models as a Friedmann-like equation. We further investigate this system by finding analytical solutions for a driven background and expand the analytical and numerical analysis for isentropic and non-isentropic perturbations. To analyse the distribution of inhomogeneities within the plasmonic system, we introduce a power spectrum, similar to the cosmological counterpart.

Here we outline the structure of the article. The next section is devoted to the presentation of the model and its analogue cosmological background expansion. We establish a Friedmann-like set of equations to the Jellium model. Section \ref{sec:perturbations} is devoted to the study of particle number perturbations in up to first order. We provide analytical solutions for well motivated cosmological like expansion rates and determine the limit of validity of this method. We also introduce the analogous power spectrum and characterize it as a function of the model parameters.
We finish the paper by summarizing the main conclusions that can be drawn up from this work in section \ref{sec:conclusao}.

\section{Hydrodynamic and cosmological picture of expanding plasmons}
\label{sec:analogue-densidade-dependente-do-tempo}

Refs. \cite{kolomeisky2019natural,kolomeisky2020normal,kolomeisky2021analog} investigate a possible analogy between Jellium systems and an expanding cosmological background. In this section, we review this problem.

Let us consider a macrocospic electron gas in the jellium model, in which particles have mass density $\rho_m = m n$ and charge density $\rho_e= e n$, where $e$ is the signed charge unit (negative for electrons), $m$ is the electron mass and $n\equiv n(t)$ the number of electrons per volume unit. In the absence of external driving forces, the gas obeys the equations \cite{ge2022comment,visser1998,kittel-quantum,kittel-intro,haas2011quantum}
	\begin{align}
	&\partial_{t} n +\vec{\nabla}(n \mathbf{v})=0 \label{eq continuidade plasmons},\\
	&m n\left(\partial_{t} \mathbf{v}+\mathbf{v} \cdot \vec{\nabla} \mathbf{v}\right)=-\vec{\nabla} p - n e \vec{\nabla} V_{e l} \label{eq de euler plasmons}, \\
    &\nabla^2  V_{el} = \frac{e}{\varepsilon} (n-n_0), \label{lei de gauss plasmons}
	\end{align}
	where $n_0$ is the fixed density of the positively charged background particles and we use $\varepsilon = 1/4\pi$ henceforth. Particle number conservation has been assumed c.f. \eqref{eq continuidade plasmons}. The dynamics is set in  \eqref{eq de euler plasmons}. Of course, gravitational effects are extremely weak in this system and have been neglected. We have also neglected any non-newtonian like behavior or extra sources of dissipation/viscosity in such system. Related to the latter, this is stated in the form of entropy conservation
    \begin{equation}\label{entropy}
        \partial_{t} S+\mathbf{v} \cdot \vec{\nabla} S=0.
    \end{equation}
In this formalism, the expanding feature of the Jellium system is well visualized, since a vanishing  electric field $\bf E$ implies $V_{el}= cte$, which in turn implies $n-n_0=0$, so a static system is possible only if it is neutral.
    
In order to find solutions for the above set of equations we fix and expanding background in which physical distance coordinates $\mathbf{x}$ are rescaled in time by the scale factor $b\equiv b(t)$ according to $\mathbf{x}=b\mathbf{q}$. By deriving the latter, the velocity can be written as $\dot{\mathbf{x}}\equiv\mathbf{v}=(\dot{b}/b) \mathbf{x}$ where $H=\dot{b}/b$ is the analogue Hubble factor and $t$ the analogue cosmic time, or the lab time. In other words, the system is experiencing the Hubble flow, analogously to what galaxies experience in cosmic expansion.  By introducing such a velocity field into \eqref{eq continuidade plasmons} one finds
\begin{equation}
    n=\frac{n_i}{(b/b_i)^3},
\end{equation}
where we have identified the particle number density at some initial time $t_i$, corresponding to a scale factor $b_i$, as $n(b_i)=n_i$. In a typical plasma discharge \cite{johnson-malter-plasma} with a partially ionized gas, one can measure the electronic density to be around $n_i \approx 2.9\times 10^9 \text{electrons}/\text{cm}^3$.

The dynamical evoltuion of such system is provided by the Euler equation \eqref{eq de euler plasmons}. Adopting the velocity field corresponding to the Hubble flow one obtains
\begin{equation}
\dot{H}+H^2=\frac{4 \pi e^2}{3m}(n-n_0).
\label{Friedmann2}
\end{equation}
Direct integration of the above equation results in 
\begin{equation}
    H^2=-\frac{8\pi e^2}{3m}\left(n+\frac{n_0}{2}\right)+\frac{2E}{m(b/b_i)^2},
\label{Friedmann1}
\end{equation}
where $E$ is an integration constant interpreted as the total energy of the system.

The correspondence with the relativistic $\Lambda$CDM cosmology is clear with the identifications $e^2\rightarrow -Gm^2$, $e^2n_0/m\rightarrow -\Lambda c^2/4\pi$ and $2E/m\rightarrow -kc^2$ where $k$ is a constant representing the spatial curvature. In the relativistic cosmological setup, it describes the global geometry of the universe: \(k=0\) (flat), \(k=+1\) (positively curved/closed), and \(k=-1\) (negatively curved/open). It is important to note that in presenting Eq.
\eqref{Friedmann1} we avoid the association of the integration constant $E$ with the particle number density as done by \cite{kolomeisky2019natural}.

Still motivated by the cosmological vocabulary, let us introduce the critical particle density parameter
\begin{equation}\label{densidade critica}
    n_c=\frac{3 H^2_im}{8\pi e^2}
\end{equation}
and rewrite \eqref{Friedmann1} as
\begin{equation}\label{Friedmannomegas}
    H^2=H^2_i\left[ -\frac{\Omega_{n}}{(b/b_i)^3}-\Omega_0+\frac{\Omega_E}{(b/b_i)^2}\right],
\end{equation}
where $\Omega's$ represent dimensionless fractional contributions of each term on the right hand side of \eqref{Friedmann1} such that
\begin{equation}
\Omega_n=\frac{n_i}{n_c}, \quad
    \Omega_0=\frac{n_0}{2n_c},\quad \Omega_E=\frac{2 E}{m H_i^2}.
\end{equation}
Since the critical density $n_c$ depends on the initial expansion rate parameter $H_i$ and the fact that \eqref{densidade critica} can be used to define the plasmon frequency as
\begin{equation}
    \omega_p^2 = \frac{3H^2_i \Omega_n}{2}=\frac{4\pi e^2 n_i}{m},
\end{equation}
this means that the existence of this critical density also provides a natural time scaling via the critical plasmon frequency $\omega_{pc}$. In order to provide a more realistic estimation for these quantities, let us assume $\Omega_n=2/3$. Then, one has $\omega_p=H_i$. Thus, by using $n_i \approx 2.9 \times 10^9$ we estimate $\omega_p =H_i \approx 3 \times 10^{9}$ Hz. 

Given that at the initial instant $H(a_i)=H_i$ the above evolution is subject to the constraint 
\begin{equation}
    1=-\Omega_n-\Omega_0+\Omega_E.
\end{equation}

Therefore, the full background evolution of the system depends on $2$ parameters with its magnitude at some arbitrary moment $t_i$ being determined by $H_i$. Since particle number is a positive defined quantity i.e., $\Omega_n>0$ and $\Omega_0>0$, one can infer from the above constraining relation that only $\Omega_E>1$ values represent physically admissible solutions, corresponding to open-like cosmological scenarios. This distinction is particularly noteworthy when contrasting the dynamics of the expanding Jellium model with the standard cosmological framework, where current observational data strongly support a spatially flat universe, corresponding to a vanishing curvature parameter.

\section{Evolution of electro-acoustic excitations}\label{sec:perturbations}

\subsection{The analogy with newtonian cosmological perturbations}

 In Ref. \cite{kolomeisky2020normal}, the evolution of overdensities in the charged distribution within the Jellium framework has been investigated, though only a limited number of solutions have been obtained. In this work, we extend the analysis to a broader class of cosmologically motivated scenarios, providing a detailed examination of all numerical solutions.

    In order to study the evolution of possible deviations from the homogeneity, let us perturb the above system of equations \eqref{eq continuidade plasmons} - \eqref{entropy}. This means that in this approximation we rewrite all physical quantities $f\equiv{\{n, \mathbf{v}, V_{el},S\}}$ as $f\rightarrow f+\hat{f}$, where the hat denotes first order quantities. 
    Clearly, there is no perturbation associated to $n_0$ i.e., $\hat{n}_0=0$, exactly as in the case of a cosmological constant $\Lambda$ which is the cosmological analogue of $n_0$.

The starting point is the linearization of the system of equations \eqref{eq continuidade plasmons}-\eqref{entropy}. This consists in collecting only the terms containing first order quantities and neglecting crossed products of two first order quantities e.g $\hat{n}^2, \hat{n}\hat{\mathbf{v}}, \hat{n} \hat{V}_{el}, \hat{\mathbf{v}}\hat{S}.$. Ref. \cite{kolomeisky2020normal} identified the full set of perturbed normal modes which consist in a superposition of two vortical, two longitudinal electroacoustic, and one entropic mode perturbation. All such modes have the exact counterpart in the newtonian perturbation theory \cite{Mukhanov:2005sc}. 

We shall focus on the scalar particle number perturbations. We remind the definition
\begin{equation}
    h=\frac{\hat{n}}{n}=\frac{\hat \rho}{\rho},
\end{equation}
with $\rho=n m$, giving rise to an equation for the evolution of the electro-acoustic modes
\begin{equation}
\ddot{h} + 2H\dot{h} +
\left(
   \frac{4\pi e^2 n}{m}
   - \frac{c_s^2}{b^2}\nabla_x^2
\right) h
= \frac{\sigma}{m n b^2}\nabla_x^2 \delta_s
\label{hperturb}
\end{equation}
In the cosmological context, $h$
 is interpreted as the matter density contrast, and Eq.\eqref{hperturb} provides the exact counterpart valid for sub‑horizon perturbations. In order to solve the above equation it becomes necessary to specify the effective fluid’s equation‑of‑state parameter. Since dark matter constitutes the dominant component driving structure formation, its equation of state plays a crucial role. Among the various possibilities, Cold Dark Matter (CDM) scenarios have emerged as the most promising candidates, effectively modeled as a pressureless fluid whose properties can be directly translated into this framework. In practise, this approximation corresponds to set $c^2_s=\sigma=0$ in \eqref{hperturb}. In Jellium however this analogy breaks down since there exists a number of theoretical and experimentally motivated equations of state for such system (see Refs. \cite{Alchagirov,Swift}). The existence of pressure solutions is the first important difference we want to highlight in such analogue description. 

If we assume the expanding Jellium model as an adiabatic process, then no heat is exchanged with the surroundings and the fully thermodynamical description is resumed to the barotropic condition $p\equiv p(\rho)$. For an ideal gas, the adiabatic condition is expressed as \cite{Hazeltine2018}
\begin{equation}
    p=w \rho^{\gamma}.
\end{equation} 
Since we have assumed the system's background expansion keeps homogeneity and isotropy along its expansion (which is equivalent to the cosmological principle statements) the pressure term does not contribute to the background evolution equations \eqref{Friedmann1} and \eqref{Friedmann2}. However, this adiabatic equation of state will contribute at first order perturbations with
\begin{equation}
    c^2_s=\frac{\partial p}{\partial \rho}=\gamma w \rho^{\gamma-1}\quad  {\rm and} \quad \delta_s=0 \, ({\rm adiabatic)}. 
\end{equation}

It is more convenient to treat seek for solution of equation \eqref{hperturb} in the Fourier space where
\begin{equation}
\label{transf fourier}
h_{k}(t)=\int d^3 r\, h({\mathbf{r}},t) e^{-i \mathbf{k.r}}.   
\end{equation} 
This allows to study the evolution of fluctuations for different wave modes $k=2\pi/\lambda$, where $\lambda$ is the characteristic perturbation  wavelength.
This effectively means $\nabla^2\rightarrow -k^2$. By adopting now the isentropic condition $\delta_s=0$ and using the scale factor as the dynamical variable Eq. \eqref{hperturb} can be rewritten as

\begin{eqnarray}\label{hperturbScaleFactor}
&&b^2h_k^{\prime\prime}+\left(3+\frac{b H^{\prime}}{H}\right)b h_k^{\prime}\\ \nonumber
&+&\left[\frac{3}{2}\frac{\Omega_n}{b^3}\left(\frac{H_i}{H}\right)^2+\left(\frac{c^2_s}{b^2}+\Sigma\frac{b}{\Omega_n}\right)\frac{k^2}{H^2}\right]h_k = 0,
\end{eqnarray}
where the symbol prime $(^{\prime})$ means a derivative with respect to the scale factor. We have rewritten the adiabatic speed of sound in terms of the reduced quantities
\begin{equation}
\label{def:velocidade do som}
c^2_s=\tilde{w}\,\left(\frac{\Omega_n}{b^3}\right)^{\gamma-1}=\frac{c^2_{si}}{b^{3(\gamma-1)}},
\end{equation}
being $c^2_{si}=\tilde{\omega} {\Omega_{n}}^{\gamma-1}$ the initial square speed of sound  and
$\tilde{w}=\gamma w (m n_c)^{\gamma-1}$. For $\gamma>1$ values the adiabatic speed of sound vanishes in the long time limit ($b\rightarrow \infty$). 


If the expansion preserves homogeneity along its evolution then entropy is expected to vanish at first order. Therefore, we assume the entropy perturbation $\delta_s$ is associated to the level of inhomogeneity given by $h$. Hence
\begin{equation}
    \frac{\sigma}{m n b^2} \nabla^2_x \delta_s \propto \frac{\Sigma b}{\Omega_n} k^2h_{k},
\end{equation}
where the dimensionless parameter $\Sigma\propto \sigma$ in \eqref{hperturbScaleFactor} quantifies the magnitude of the entropic modes. The isentropic case is recovered with $\Sigma=0$.

\subsection{Controlled adiabatic scenarios: Exact solutions }

Now, we start to seek for exact solutions of Eq. \eqref{hperturb}, or equivalently \eqref{hperturbScaleFactor}. In the special case where the perturbations are isentropic ($\sigma=0$) the Fourier transform of the resulting evolution equation \eqref{hperturb} becomes
\begin{equation}\label{eq fourier}
\ddot{h}_k + 2H \dot{h}_k + \left[\frac{\omega_p^2}{b^3}+\frac{c_s^2 k^2}{b^2} \right]h_k = 0.
\end{equation}

The equation above does not admit analytical solutions for the general form of $H$ given by \eqref{Friedmannomegas}. For such case, we will treat the equation numerically and present the results in the next subsection. 

For now, we shall abandon the definition of $H$ in \eqref{Friedmannomegas} and instead fix the evolution of the scale factor, consequently fixing the expansion rate $H$. This is indeed a deliberate choice, because the analogy with cosmology provides us with some expansions of theoretical interest.  It is important to note that these expansion factors cannot be obtained from the free dynamical equation presented in Eq. \eqref{eq de euler plasmons}. However, it is possible to add to Eq. \eqref{eq de euler plasmons} a well-defined external force $F\mathbf{r}$  that modifies $n_0$ in the analogue Friedmann equation \eqref{Friedmann1} as $n_0 \to n_0 - F$, creating a free parameter that can be manipulated yielding a specific desired expansion. Thus, in the ideal case of well-controlled laboratory conditions any desired expansion can be achieved. Also, given the background nature of this external field, it would not possess perturbations backreacting into the particle number fluctuations and hence Eq. \eqref{eq fourier} would remain valid. We provide below analytical solutions for two well motivated  cosmological background cases.

$\bullet \ b(t) \propto t^{2/3}$: The expansion rate behaves as $H = 2/3t$ regardless of the pressure of the fluid. In the analogue cosmological case this represents the matter dominated phase which lasts $\sim 10 $Gys, being important for the structure formation process. However, in the cosmological case this expansion is achieved only in a fully pressureless dominated fluid. Here, we have the freedom to set $b(t) \propto t^{2/3}$ even with $\omega \neq 0$. Hence,  equation \eqref{hperturb} can also be written
as
\begin{eqnarray}
\label{fnpeg3}
\ddot h_k + \frac{4}{3t}\dot h_k + \biggr\{\frac{\omega_p^2}{t^2}+\frac{ c_{si}^2k^2}{t^{\frac{2(3\gamma -1)}{3}}}\biggl\} h_k = 0,
\end{eqnarray}
Equation (\ref{fnpeg3}) can be solved using two different transformations.
Defining the new variables,
\begin{eqnarray}
y\equiv k c_{si}t ^r, \quad r \equiv \frac{4}{3} - \gamma,
\end{eqnarray}
it turns into,
\begin{eqnarray}\label{ddeltadx}
\frac{d^2 h_k}{dy^2}+ \left(1+\frac{1}{3r}\right)\frac{1}{y}\frac{d h_k}{dy} + \biggr\{1 + \frac{\omega_p^2}{r^2y^2}\biggl\}h_k = 0.
\end{eqnarray}
Now, we redefine the function $h_k$ as,
\begin{eqnarray}
h_k = y^{-\frac{1}{6r}}\lambda_k.
\end{eqnarray}
In terms of the variable $\lambda_k$ equation \eqref{ddeltadx} turns into,
\begin{eqnarray}
\frac{d^2\lambda_k}{dy^2} + \frac{1}{y}\frac{d \lambda_k}{dy} + \biggr\{1 + \biggr(\frac{5 \omega_p}{6r}\biggl)^2\frac{1}{y^2}\biggl\}\lambda_k = 0.
\end{eqnarray}
The solution, restoring the variable $t$, is
\begin{equation}
h_k = t^{-1/6}\biggr\{\mathcal{A}J_{5\omega_pi/6r}(kc_{si} t^r) + \mathcal{B}J_{-5\omega_pi/6r}(kc_{si} t^r)\biggl\} .
\end{equation}
where $\mathcal{A}$ and $\mathcal{B}$ are constants. The solution is given in terms of the Bessel functions $J_\nu(kc_{s0}t^r)$, where $\nu$ is order of the Bessel function (imaginary in this case) and $kc_{si} t^r$ its argument.


$\bullet \ b(t)\propto e^{H_i t}$: A simple inflation-like scenario is modeled using a de Sitter-like expansion factor $b(t)\propto e^{H_i t}$. We introduce the variable $z = b^{-1}$ and use the transformation $h_k(t(z)) = \mathcal{J}(z)$, to reveal 
\begin{equation}
\label{eq funcao J}
    \frac{d^2\mathcal{J}}{dz^2} - \frac{1}{z}\frac{d\mathcal{J}}{dz} + \left( \frac{c_{si}^2 k^2}{H_i^2} z^2 + \frac{\omega_p^2}{H_i^2}z \right)\mathcal{J} = 0,
\end{equation}
where we focus on the special case $\gamma = 5/3$ for a non-relativistic ideal electron gas at temperature $T > 0$. Using $x = ({\omega_p^2}/{H_i^2})^{1/3} \ z$, the regular solution at the origin is calculated as
\begin{multline}
\label{frobenius funcao J}
    \mathcal{J}(x) = x^\alpha \exp\left[\frac{i}{2}\left(\beta x + \zeta x^2\right)\right]
    \\
    \times\mathrm{HeunB}\left[(2\alpha - 1)\beta, \beta^2 - 4\alpha\zeta,2 \alpha -1,2 \beta ,4 \zeta, x\right].
\end{multline}
In principle, we may choose any combination of $\alpha,\beta,\zeta$ within the following list of options
\begin{equation}
    \alpha = \{0,2\}, 
    \quad 
    \beta = \pm\frac{1}{g_0}, 
    \quad
    \zeta = \pm g_0
\end{equation}
where $g_0$ is
\begin{equation}
         g_0^2 = \left(\frac{c_{si} k}{\omega _p}\right)^2 \sqrt[3]{\frac{H_i}{\omega _p}}.
\end{equation}
Due to the properties \cite{ronveaux, black-string} of the biconfluent Heun function, the quantity $2\alpha - 1$ cannot be a negative integer, and so we choose $\alpha = 2$ and the positive branches $\beta=+1/g_0$ and $\zeta=+g_0$, obtaining a regular solution at early and late times.
In the limit $g_0\to 0$, or equivalently, in the limit where $\omega_p$ is dominant,
\begin{equation}
    \mathcal{J}(x) \to c_1 \mathrm{Ai}'\left(- x\right)+c_2 \mathrm{Bi}'\left(-
   x\right)
\end{equation}
which are the derivatives of the Airy functions, closely related to Bessel functions \cite{NIST:DLMF}. 
\begin{figure}[h]
    \centering
    \includegraphics[width=.9\linewidth]{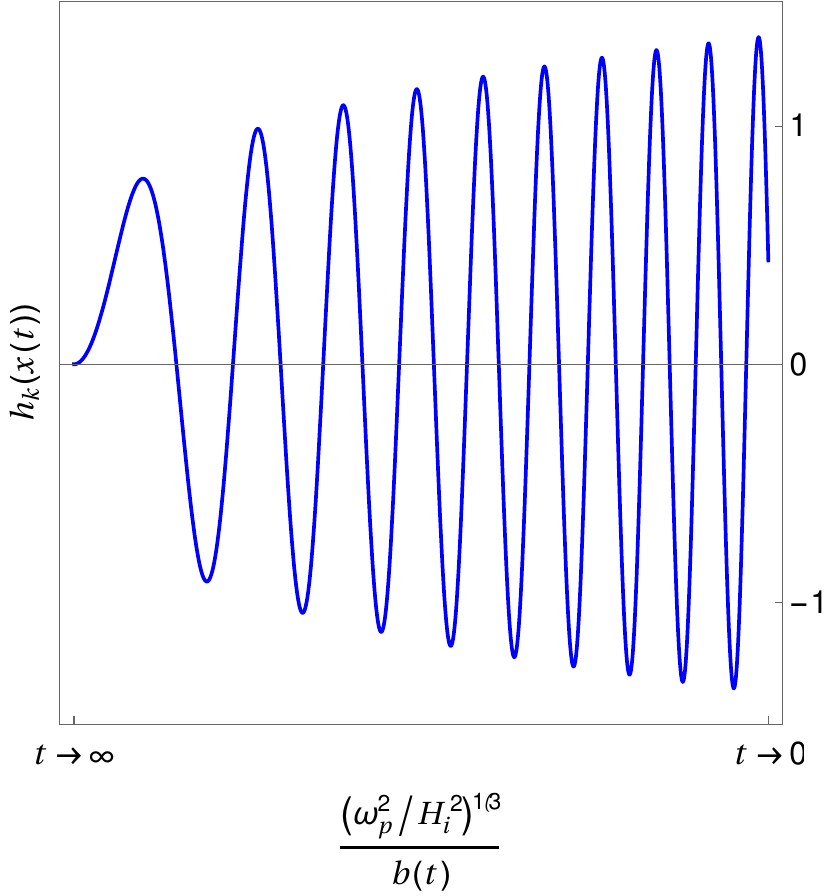}
    \caption{Evolution of the frequency-dominated amplitudes $h_k$ for a de Sitter like fixed background expansion.}
    \label{fig:airy}
\end{figure}
We see that, imposing the asymptotic condition $\mathcal{J}(0) = 0$, the amplitude $h_k$ vanishes as $t\to\infty$. This result is expected, since the exponential expansion of the background will render the fluctuations negligible in the future asymptotic limit $t\rightarrow \infty$.
\subsection{Numerical solutions}

In order to provide numerical solutions for $h_k$ from its initial conditions we set that the system starts to evolve at a initial time $t_i=0$ (which corresponds to a unitary scale factor $b_i=b(t_i)=1$) at an expansion rate $H(t_i)=H_i=1$. So as to obtain the evolution of the perturbations $h_{k}$ we solve numerically \eqref{hperturbScaleFactor}. We set the initial conditions as follows. With the initial condition $h_k(b_i)$ we set the inhomogeneity level of the charge distribution when the experiment starts to take place. Let us say the system starts with a $1\%$ inhomogeneity level i.e., $h_k(b_i)=0.01$. This choice is arbitrary and depends on the experimental apparatus. Also, there is no physical mechanism to set the same initial perturbed amplitude for all scales. Contrarily to the cosmological scenario where a Harrison-Zeldovich primordial spectrum (see definition below in \eqref{Pk}) is well justified \cite{Harrison,Zeldovich}. We also set $h^{\prime}_k(b_i)=1$ assuming the growing mode dominates initially as already stated by Ref. \cite{kolomeisky2020normal}.  

The so called Coulomb explosion is a particular case with $\Omega_0=0$. The parameter $\Omega_0$ influences the evolution of perturbations only indirectly via its influence on the background expansion. Remind that $\hat{n}_0=0$. Hence, $\Omega_0\neq0$ values will have only a negligible impact on the qualitative aspects of the perturbative analysis. We adopt therefore $\Omega_0=0$ henceforth.

A number of possible perturbative scenarios can be studied from this point on.  

A) Long-wavelength limit: Ref. \cite{kolomeisky2020normal} analysed \eqref{hperturb} in this limit which is technically equivalent to pressureless cases. In this limit the wave-number dependence is eliminated i.e., $h\equiv{h(t)}$. Hence, by setting $c^2_s=\sigma=0$, the general solution of Eq. \eqref{hperturb} is written in the form $h(t)=A \,h_1(t) +B\, h_2(t)$ where $A$ and $B$ are determined by initial conditions. The two independent solutions can be easily found as 
\begin{equation}
    h_1 \propto H(t),\quad h_2(t)\propto H\int^{b} \frac{db}{\dot{b}^3}.
\end{equation}

Adopting the scale factor as the dynamical variable, the explicitly expressions for the Coulomb explosion case can be obtained in terms of the free model parameter $\Omega_n$. Such solutions are identified as the decaying mode
\begin{equation}
    h_1 \propto \frac{\sqrt{b + (b-1)\Omega_n}}{b^{3/2}},
\end{equation}
and the asymptotic growing-to-satured mode
\begin{multline}\label{h2}
h_2 \propto \ \frac{b - 3\Omega_n + b\Omega_n}{b(1 + \Omega_n)^2}
\\ 
+ \frac{3\Omega_n \sqrt{b - \Omega_n + b\Omega_n} \,
\operatorname{arctanh}\!\left(
   \frac{\sqrt{b - \Omega_n + b\Omega_n}}{\sqrt{b}\sqrt{1 + \Omega_n}}
\right)}{b^{3/2}(1 + \Omega_n)^{5/2}}.
\end{multline}
The latter will be better explored in the subsequent analysis.

The perturbative formalism developed so far is restricted to the so‑called linear regime, i.e., the magnitude of the perturbations is assumed to be small compared to their background reference value. This implies 
$h_k<<1$. However, inhomogeneities evolve over time. For the growing modes, they amplify in magnitude and can eventually reach values $h_k \sim 1$. To investigate this regime, the nonlinear terms previously discarded must now be retained. With this procedure, the nonlinear version of \eqref{hperturbScaleFactor} reads

\begin{eqnarray}\label{nonlinearhperturbScaleFactor}
&&b^2h_k^{\prime\prime}+\left(3+\frac{b H^{\prime}}{H}\right)b h_{k}^{\prime} -\frac{4b^2}{3}\frac{(h_{k}^{\prime})^2}{(1+h_{k})}\\ \nonumber
&+&\left[\frac{3}{2}\frac{\Omega_n}{b^3}\left(\frac{H_i}{H}\right)^2+\left(\frac{c^2_s}{b^2}+\Sigma\frac{b}{\Omega_n}\right)\frac{k^2}{H^2}\right]h_k(1+h_{k}) = 0,
\end{eqnarray}

\begin{figure} [t]
    \centering    \includegraphics[width=0.9\linewidth]{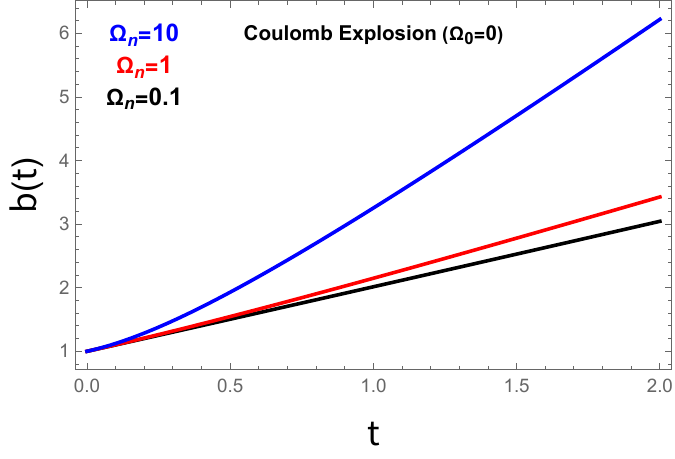}
\includegraphics[width=0.9\linewidth]{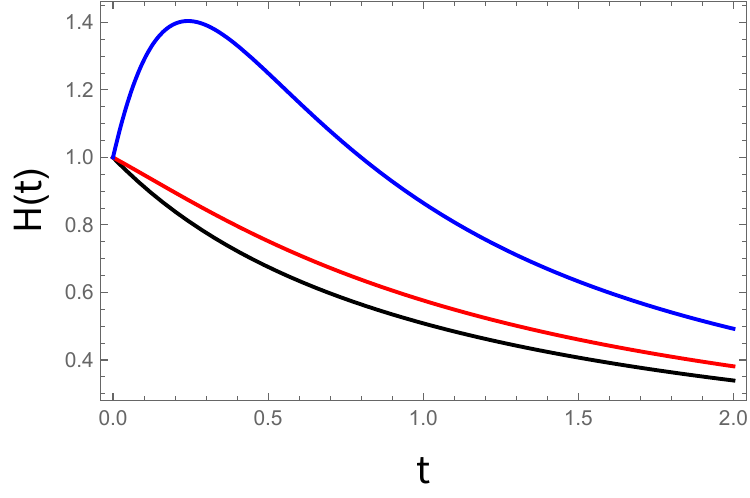}   \includegraphics[width=0.9\linewidth]{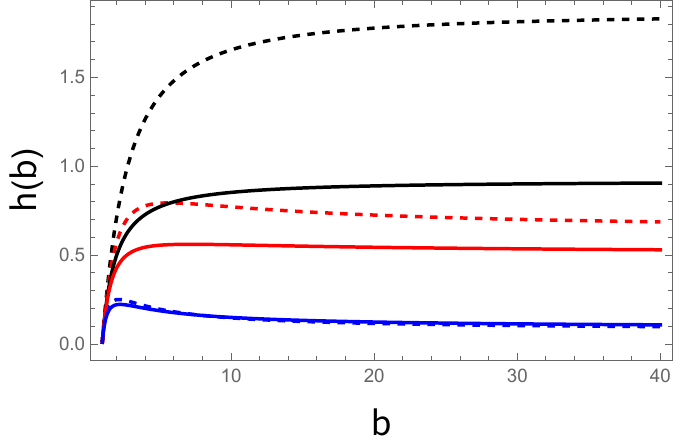}
   \caption{Coulomb explosion scenarios $\Omega_0=0$ in the Long-wavelength limit. The dynamical evolution for the scale factor (top), the expansion rate (middle) and the excitation clustering (bottom) are displayed for different $\Omega_n$ values according to the color scheme.}
    \label{CoulombExplosion}
\end{figure}

We shown in Fig. \eqref{CoulombExplosion} the evolution of the scale factor (top panel) and the expansion (middle panel) as a function of the lab time in arbitrary units. In the bottom panel of this figure we provide the numerical solution for $h_k$ as a function of the scale factor for the linear case \eqref{hperturbScaleFactor} (solid lines) and the non-linear one \eqref{nonlinearhperturbScaleFactor} (dashed). The adopted charge density parameter values follow the color scheme shown in the top panel inset i.e. $\Omega_n=0.1$ (black), $\Omega_n=1$ (red) and  $\Omega_n=10$ (blue). Clearly, the faster the expansion rate, the weaker the clustering evolution becomes. All solutions for $h$ tend to a constant value in the long-time limit (according to the solution found in \eqref{h2}).  The magnitude of the parameter $\Omega_{n}$ enhances the expansion rate, which in turn suppresses the growth of electroacoustic perturbations. In other words, the more dilute the system is, the larger the amplitude of the perturbations. Once more, it is also important to emphasize that the perturbative expansion employed to obtain \eqref{hperturb} remains valid as long as the condition $h<1$ is fulfilled. Then, according to these plots, given the initial $1\%$ inhomogeneity level $h(b_i)=0.01$ the linear formalism breaks down for dilute systems characterized by $\Omega_n \lesssim 0.1$ (black solid curve)  since $h\sim 1$ values have been reached. Hence, the full nonlinear corrections should be included via \eqref{nonlinearhperturbScaleFactor}. The dotted curves in the bottom panel refer to the full solution where the nonlinear corrections are taken into account. Otherwise, for higher densities (blue curves) the linear regime remains valid. Indeed, note that the solid and dotted blue curves are almost indistinguishable.

B) Short-wavelength limit: For all $c^2_s \neq 0$ or $\Sigma \neq 0$ cases the scale dependence is evident. Hence, in this case the electro-acoustic amplitude depend on the wavenumber $h_k \equiv h(k,t)$. Both isentropic scenarios ($w\neq 0$ and $\Sigma=0$) as well as non-isentropic scenarios ($\Sigma \neq 0$) can also be studied.   

\begin{figure}[ht]
    \centering    \includegraphics[width=0.9\linewidth]{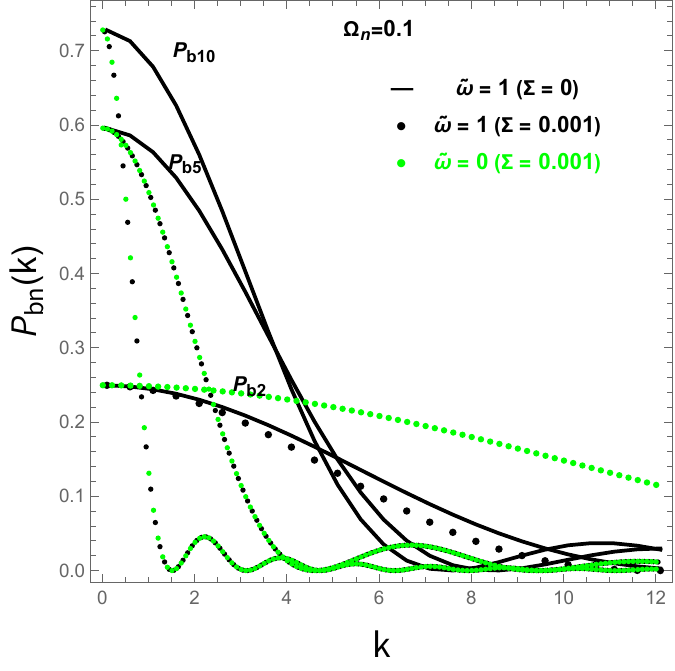}
   \caption{Number particle density clustering power spectrum $P_{bn}(k)=|h(k,b= n a_i)|^2$ as defined in \eqref{Pk} for the Coulomb explosion scenario. The solid lines follow isentropic modes with $\Sigma=0$. The dots show the departure from the isentropic case with $\Sigma=0.001$.  }
    \label{PowerSpectrum}
\end{figure}

Inspired again by the analogy with the cosmological setup and also assuming the perturbations are distributed according to a homogeneous Gaussian random process, let us define the quantity $P_{bn}$ which represents the power spectrum evaluated when the system scale factor $b$ reaches $n$ times its initial value $b_i$,
\begin{equation}\label{Pk}
    P_{bn}(k)=|h_k(k,b= n b_i)|^2.
\end{equation}
This definition provides a powerful tool to connect theory with experimental results. We show in Fig. \ref{PowerSpectrum} the power spectrum for different parameter space configurations. Solid black lines consider an isentropic ($\Sigma =0$) equation of state parameter $\tilde{w}=1$ with $\gamma=5/3$ evaluated when the system reaches twice its initial size $b=2b_i$ $(P_{b2})$ and also when it reaches five and ten times the original size. According to definition \eqref{def:velocidade do som}, this corresponds to an initial square speed of sound $c^2_{si}=0.215$. Given the arbitrary units adopted here, in the wavenumber regime ($k \lesssim 4$) the power spectrum amplitude grows with the time scale. In the long wavenumber limit $P_{bn}$ tends to zero in all scenarios due to the damping effect caused by the expansion rate. Now, by keeping $\tilde{w}=1$ and switching on the entropic parameter $\Sigma$ the inhomogeneity is suppressed as shown by the black dots. The non-isentropic ($\Sigma=0.001$) pressureless ($\tilde{w}=0$) power spectrum follows the green dots. Initially, at $b=2b_i$, pressureless systems experiences a faster departure from homogeneity (green dots) when compared to the pressure supported scenarios (black dots), but both tend to the isentropic scenarios at later times. This result strongly suggests the existence of an optimal timescale for collecting experimental data that characterizes the system’s features.

\section{Discussion and final remarks}\label{sec:conclusao}
We investigated a 3D Jellium as a toy model to explore analogue cosmological scenarios. In this three-dimensional homogeneuous and isotropic model, the system's velocity field follows the Hubble flow, presenting therefore a Friedmann-like background evolution. According to \eqref{Friedmannomegas} this scenario naturally present the same functional form as an anti-de Sitter spacetime with a negative pressure contribution with curvature. The Jellium particle number perturbations present a similar structure to the perturbations in the newtonian cosmological fluid, except that because of the mapping $-Gm^2 \to e^2$, the system does not present an analogue for the Jeans instability \cite{kolomeisky2019natural}. To illustrate cosmologically relevant scenarios, we selected two expansion factors, $b(t) \propto t^{2/3}$ and $b(t) \propto e^{H_i t}$, for which the amplitude of our perturbations, described by \eqref{eq fourier} in the isentropic case, have exact solutions, both decaying as $t\to \infty$. With the numerical evaluation provided in the bottom panel of Fig. \ref{CoulombExplosion} we can determine the limit of validity of the linear regime in the perturbative expansion as a function of the density parameter $\Omega_n$.

We have introduced the power spectrum of fluctuations, $P_{bn}(k)$, which constitutes a central diagnostic tool for connecting theoretical predictions with experimental realizations. Its temporal evolution and dependence on the wavenumber have been illustrated in Fig. \ref{PowerSpectrum} across a broad range of physical scenarios. In particular, the role of non‑isentropic contributions has been highlighted, revealing their significant impact on the spectral features.

Beyond this analysis, several cosmological aspects remain open to exploration. These include the mechanisms of analogue particle production, the emergence of large‑scale structure within the Jellium framework, and the propagation of background inhomogeneities. Each of these directions offers the potential to deepen the analogy between condensed‑matter systems and cosmological dynamics. Moreover, ongoing efforts are directed toward experimental measurements of physical observables in Jellium, conceived as analogues of cosmological quantities, thereby paving the way for a more direct confrontation between theory and experiment. \textcolor{black}{Considering that up to this point most of the efforts from the analogue gravity community were focused on experiments based on Bose-Einstein condensates, the Jellium model presents itself as a natural and arguably more feasible alternative to test basic features of early universe cosmological models.}

%
\begin{acknowledgments}
M. E. P. thanks Dr. Carla R. Almeida, Dr. Rodrigo Gonçalves, Diego T. P. Silva, Dra. Débora C. M. Rodrigues, Dr. Licínio Portugal and Dr. Ge Li for the stimulating exchange of ideas. M. E. P. and A. G. M. S. gratefully acknowledge CNPq (grant numbers 140471/2022-7 and 309052/2023-8) for partial financial support. HV acknowledges CNPq (grant number 311389/2023-6).  FBL is funded by Fundação Cearense de Apoio ao Desenvolvimento Científico e Tecnológico (FUNCAP) and by  Conselho Nacional de Desenvolvimento Científico e Tecnológico (CNPq), grant number 305947/2024-9.
\end{acknowledgments}
%
%
\bibliography{apssamp}
\end{document}